# Modeling Fibril Fragmentation in Real-Time

Pengzhen Tan, Liu Hong[*]

Zhou-Pei Yuan Center for Applied Mathematics, Tsinghua University, Beijing, 100084, P.R. China

[*]zcamhl@tsinghua.edu.cn

**ABSTRACT** During the application of mass-action equation models to the study of amyloid fiber formation, time-consuming numerical calculations constitute a major bottleneck when no analytical solution is available. To conquer this difficulty, here an alternative efficient method is introduced for the fragmentation-only model. It includes two basic steps: (1) simulate close-formed time-evolution equations for the number concentration $P(t)$ derived from the moment-closure method; (2) reconstruct the detailed fiber length distribution based on the knowledge of moments obtained in the first step. Compared to direct solution, current method speeds up the calculation by at least ten thousand times. The accuracy is also quite satastifactory if suitable forms of approximate distribution fucntion is taken. Further application to $PI_{264}$-$b$-$PFS_{48}$ micelles study performed by Guerin *et al.* confirms our method is very promising for the real-time analysis of the experimental data on fiber fragmentation.

## INTRODUCTION

The linear aggregation of soluble peptides and proteins into insoluble amyloid fibrils is a typical self-assembling phenomenon of bio-macromolecules[1], and is closely related to many well-known human neurodegenerative diseases[2]. Thus to uncover the mechanisms of amyloid fiber formation, will not only have great theoretical values, but also shed light on the medical diagnosis and treatment of amyloidosis[3].

In the past decades, related to this hot topic, many efforts have been dedicated to quantify the effects of primary nucleation and elongation[4,5]. While fragmentation[6] and other processes, like fiber surface facilitated nucleation[7], lateral thickening[8] and *etc.*, have received far less attention.

However as we know, single filament becomes mechanically unstable and tend to break when its length exceeds certain values[9]. Even for bundled fibrils in vivo, fragmentation is unavoidable in the presence of mechanical stress, thermal motion, or chaperones such as Hsp104, which has a known ability to fragment fibril samples[10]. Actually as a special kind of secondary nucleation, fragmentation can effectively accelerate the fiber formation process by providing new seeds[11], affect the scaling relations between kinetic quantities (like the lag-time and maximum fiber growth rate) and protein concentration (from critical-nucleus-size dependent to independent)[12], alter the detailed fiber length distribution from exponentially decaying to bell-shape-like[13], and even enhance the toxicity of fibril samples to disrupt membranes and to reduce cell viability[14].

To quantify the key role of fragmentation played during the formation of breakable amyloid fiber, various experiments, like the shear flow[15,16] and sonication studies[17,18], are designed. However to interpret the experimental results, especially to provide a quantitative relationship between the observed data and their underlying mechanisms, is not an easy task. Traditional way is to simulate the kinetic models formulated in a form of mass-action equations, from which very detailed time-evolutionary information about fiber length distribution can be obtained. But this approach suffers from intrinsic calculational difficulties due to the high-dimensional nature of the ODE system and cannot meet the requirement of real-time analysis in real applications.

In current paper, to solve above problems, an alternative efficient method is introduced, which is constituted by two successive steps (see Fig. 1): (1) calculate the time-evolutionary data of statistical moments, such as the number concentration and mass concentration of fibrils, through moment-closure equations; (2) reconstruct the detailed fiber length distribution based on the knowledge of moments obtained in the first step. A most outstanding advantage of the current method lies in its efficiency. Compared to direct simulation of original mass-action equations, our method speeds up the calculation by at least ten-thousand times (from days to seconds). Further counting the tedious data fitting procedure, the performance will be far more striking. The accuracy of my method is also quite satastifactory if suitable forms of approximate distribution fucntion is taken. These two features guarantee our method as a promising solution to the real-time analysis of the experimental data on fiber fragmentation.

In accordance with the procedure of our method, we will firstly make a brief introduction to the fragmentation-only model. Then the central model -- close-formed time-evolution equations for the number concentration $P(t)$ are derived based on the moment-closure method. By the knowledge of moments, two different empirical formulas are proposed to rebuild the detailed fiber length distribution with quantitative comparisons. Finally we apply the current method to the $PI_{264}$-$b$-$PFS_{48}$ micelles study performed by Guerin *et al.*[17].

## MODEL & METHODS

### Mass-action equations

To analyze the effect of fragmentation, let's consider a most simple model, which includes only two processes – fibril fragmentation and association[19],

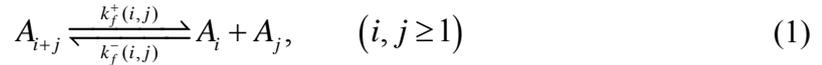

$$A_{i+j} \underset{k_f^-(i,j)}{\overset{k_f^+(i,j)}{\rightleftharpoons}} A_i + A_j, \qquad (i, j \geq 1) \tag{1}$$

where $A_i$ stands for fibrils in size $i$. The length-dependent reaction rate constants for fibril fragmentation $k_f^+(i,j)$ and association $k_f^-(i,j)$ are chosen according to the Gaussian model introduced in Ref. [16] with some minor modifications, *i.e.*

$$\begin{cases} k_f^+(i,j) = k_f^+ \dfrac{(i+j)^{x-1}}{\sqrt{2\pi} R} \exp\left[-\dfrac{(i-j)^2}{8(i+j)^2 R^2}\right], \\ k_f^-(i,j) = 0. \end{cases} \tag{2}$$

Then based on the law of mass action, the time-evolution equations for $[A_i]$ -- the molar concentration of fibrils in size $i$ can be written as

$$\begin{aligned}\dfrac{d}{dt}[A_i] &= 2\sum_{j=i+1}^{\infty} k_f^+(i, j-i)[A_j] - \sum_{j=1}^{i-1} k_f^+(j, i-j)[A_i] \\ &\quad -2\sum_{j=1}^{\infty} k_f^-(i,j)[A_i][A_j] + \sum_{j=1}^{i-1} k_f^-(j, i-j)[A_j][A_{i-j}]. \end{aligned} \tag{3}$$

Thanks to the conservation law of mass, we also have $\sum_{i=1}^{\infty} i \cdot [A_i] = m_{tot}$.

As we have pointed out, due to the high-dimensional nature of the ODE system involved ($\max\{i\} = 10^3 - 10^6$), direct simulation of above equations will be extremely time-consuming (at least in days) and cannot meet the requirements of real-time analysis in application. Therefore some more efficient approaches should be considered. This will constitute the major task of our following paragraphs.

**Moment-closure method**

It is easy to check that for above choice of $k_f^+(i,j)$ and $k_f^-(i,j)$ in Eq. 2 (actually for general length-dependent rates), no closed time-evolution equation for $P(t) = \sum_{i=1}^{\infty}[A_i]$ can be derived from Eq. 3 (note $M(t) = \sum_{i=1}^{\infty} i \cdot [A_i]$ is conserved). To solve this problem, we adopt the moment-closure method introduced in our previous studies[20]. For this purpose, we define the free energy function $F$ of the system as

$$F/(k_B T) = \varepsilon_f \sum_{i=1}^{\infty}[A_i] + \varepsilon_m \sum_{i=1}^{\infty} i \cdot [A_i] \\ + \sum_{i=1}^{\infty}([A_i]\ln[A_i] - [A_i]) - (x-1) \cdot \sum_{i=1}^{\infty} \ln i \cdot [A_i], \quad (4)$$

where $k_B$ is the Boltzmann constant and $T$ is the temperature. $\varepsilon_f > 0$ represents the free energy penalty associated with the boundary region of a fiber, $\varepsilon_e < 0$ captures the averaged monomeric free energy gained from various interactions and conformational constraints, such as hydrogen bonds, hydrophobic interactions, side-chain packing *etc*. The third term in Eq. 4 comes from the mixing entropy of different fibril species as in the Flory-Huggins theory[21], while the last one accounts for the entropic contribution originated from internal freedoms (like translation and rotation) of a fiber moving in the water solution.

With this, we consider the minimization of the free energy under proper constraints,

$$\min \quad F([A_i]) \quad (5)$$

$$\text{s.t.} \quad \sum_{i=1}^{\infty}[A_i] = P, \quad \sum_{i=1}^{\infty} i \cdot [A_i] = M. \quad (6)$$

The two constraints in Eq. 6 come from the targeted number concentration $P(t)$ and mass concentration of fibrils $M(t)$.

Because the free energy is convex with respect to the distribution $([A_1],[A_2],[A_3],\cdots)$, above constrained optimization problem can be solved simply by taking the variation

$$\frac{\delta}{\delta[A_i]}\left[F([A_i])/(k_B T) - \lambda_1\left(\sum_{i=1}^{\infty}[A_i] - P\right) - \lambda_2\left(\sum_{i=1}^{\infty} i \cdot [A_i] - M\right)\right] = 0, \quad (7)$$

with Lagrangian multipliers $\lambda_1$ and $\lambda_2$ corresponding to the two constraints. It directly follows that

$$[A_i] = i^{x-1} \cdot \exp\left[\lambda_1 + \varepsilon_f + i(\lambda_2 + \varepsilon_m)\right], \quad (8)$$

where the unknown parameters $\lambda_1 + \varepsilon_f$ and $\lambda_2 + \varepsilon_m$ can be determined by

$$\begin{cases} \sum_{i=1}^{\infty} i^{x-1} \cdot \exp\left[\lambda_1 + \varepsilon_f + i(\lambda_2 + \varepsilon_m)\right] = P, \\ \sum_{i=1}^{\infty} i^{x} \cdot \exp\left[\lambda_1 + \varepsilon_f + i(\lambda_2 + \varepsilon_m)\right] = M. \end{cases} \quad (9)$$

**Moment-closure equations**

Substituting the distribution function in Eq. 8 into original mass-action equations (Eq. 3), simple moment-closure equations for the time evolution of $P(t)$ can be derived, *i.e.*

$$\begin{cases} \dfrac{dP}{dt} = k_f^+ \dfrac{\alpha}{\sqrt{2\pi}R} \Xi(\theta; x, R), \\ \alpha \Sigma(\theta; x-1) = P, \\ \alpha \Sigma(\theta; x) = M, \end{cases} \quad (10)$$

where $\alpha = \exp(\lambda_1 + \varepsilon_f)$, $\theta = \exp(\lambda_2 + \varepsilon_m)$, and $M(t) = m_{tot}$ due to the mass concervation law. The empirical functions $\Xi(\theta; x, R) = \sum_{i=1}^{\infty} \sum_{j=1}^{\infty} (i+j)^{2x-2} \exp\left[-\dfrac{(i-j)^2}{8(i+j)^2 R^2}\right] \theta^{i+j} \sim \Xi_0(x) \theta^2 (1-\theta)^{-2x}$ with $\Xi_0(x) \approx 0.5 + \exp(2.1x - 3.1) + \exp(0.19x^2 + 2.6x - 4.98)$ (See Figs. 2A-2D), and $\Sigma(\theta; x) = \sum_{i=1}^{\infty} i^x \theta^i \sim \Sigma_0(x) \theta (1-\theta)^{-(x+1)}$ with $\Sigma_0(x) \approx 0.35 + \exp(0.092x^2 + 0.77x - 1.38)$ (see Figs. 2E and 2F) are found through extensive numerical experiments, providing $x \geq 1, R \geq 1$.

Actually from the moment-closure equations, not only $P(t)$ but also other high-order moments could be solved[22]. As shown in Fig. 3, solutions of above equations (Eq. 10) match perfectly with numerical solutions of original mass-action equations (Eq. 3), except for a very short time region at the beginning. This phenomenon has been understood as "initial layer" (some kind of boundary effect) in the literature. Interested readers can refer to [23].

**Reconstruct Fiber Length Distribution**

The last and most crucial step is to reconstruct an accurate fiber length distribution based on the knowledge we have learned from the moment-closure equations. A natural choice is the distribution function $[A_i] = \alpha i^{x-1} \theta^i$ in Eq. 8 with constraints $\sum_{i=1}^{\infty} [A_i] = P$, $\sum_{i=1}^{\infty} i \cdot [A_i] = M$. The fiber length distribution thus derived is illustrated by green dotted curves in Fig. 4. It is found that although the correct behaviors of fiber length distribution are reproduced, quantitative comparisons with the exact one obtained from mass-action equations (red solid curves) are still not very satisfactory.

Nevertheless if more realistic distribution function as well as additional information about high-order moments is considered, much better results can be achieved. For this purpose, we assume the fiber length distribution obeys a modified Gaussian form

$$[A_i] = \alpha i^{x-3} \exp\left[-(i-\bar{i})^2/\sigma^2\right], \tag{11}$$

in which the three fitting parameters $\alpha$, $\bar{i}$ and $\sigma$ can be uniquely determined from following constraints $\sum_{i=1}^{\infty}[A_i] = P$, $\sum_{i=1}^{\infty} i \cdot [A_i] = M$ and $\sum_{i=1}^{\infty} i^2 \cdot [A_i] = ML_w$. From the blue dashed curves in Fig. 4, a much better agreement can be seen, which directly confirms the validity of our method.

**DISCUSSIONS**

In many different areas, including chemical reactions, combustion, system biology and so on, models formulated in mass-action equations are widely used. However due to the great complexity of real problems, the ODE system thus constructed is generally very large and requires extremely long time for calculation. This constitutes a major bottleneck in practice. Thus to simplify the calculation of mass-action equations not only has great theoretical interest, but also is essential for real applications. This provides the basic motivation of our current study.

To solve above problem, here we introduce an efficient approach, which separates the calculation into two successive steps: (1) calculate the time-evolutionary data of statistical moments by moment-closure equations; (2) reconstruct the detailed fiber length distribution based on the knowledge of moments obtained in step one. As illustrated by the fragmentation-only model, our method speeds up the calculation by at least ten-thousand times (from days to seconds), and can act as a promising solution to the real-time analysis of the experimental data on fiber fragmentation.

Not limited to the Gaussian model, even when completely different reaction rate constants for fiber fragmentation and association are taken, our method still works very well (see Hill's model in the Supporting Materials). Moreover in recent studies, the same idea has been applied to more complicated models with nucleation, elongation and fragmentation[24], which shows our method has a much broader applicable range.

Another notable point is related to the calculation of functions $\Xi(\theta; x, R)$ and $\Sigma(\theta; x)$ in Eq. 10, which will directly affect the efficiency and accuracy of the moment-closure equations. Here to achieve a high speed, we take the advantage of $\theta \to 1$ in most calculations and construct empirical formulas through extensive numerical experiments. In most cases ($x \geq 1, R \geq 1$), this approach really works. But for $0 \leq x < 1$ or $R \ll 1$, further modifications should be taken. A much safer suggestion in practice is to make a reference table for functions $\Xi(\theta; x, R)$ and $\Sigma(\theta; x)$ with specific $x$ and $R$ before running simulation.

**ACKNOWLEDGEMENTS**
The authors thank the helpful discussions of Prof. Wen-An Yong. This work was partially supported by the National Natural Science Foundation of China(NSFC 11204150) and by Tsinghua University Initiative Scientific Research Program (20121087902).

principle. Here we choose the zeroth-order moment $P(t)$ and the first-order moment $M(t)$ as constraints. Consequently, moments in second order and above can be expressed as a function (generally implicit) of $P(t)$ and $M(t)$. This fact will be used in the calculation of weight-average length $L_w(t)$ in Fig. 3A.

23. Yong WA (1999) Singular perturbations of first-order hyperbolic systems with stiff source terms. *J. Differ. Eq.* 155:89-132.
24. Hong L (2012) Link the fluorescence and TEM studies on amyloid fiber formation -- To reconstruct the fiber length distribution based on the knowledge of moments evolution. arXiv:1212.4555.

**FIGURE LEGEND**

Fig. 1 A flowchart for our approach.

Fig. 2 Asympotic behaviors of functions (A-D) $\Xi(\theta;x,R)$ and (E-F) $\Sigma(\theta;x)$. Two significant scaling regions can be found, *i.e.* when $\theta \to 0$, $\Xi(\theta;x,R) \propto \theta^2$, $\Sigma(\theta;x) \propto \theta$; when $\theta \to 1$, $\Xi(\theta;x,R) \propto (1-\theta)^{-2x}$, $\Sigma(\theta;x) \propto (1-\theta)^{-(x+1)}$ (the same conclusions could be reached for other cases with $x \geq 1$). Another notable point in plots B and D is that $\Xi(\theta;x,R)$ gets indepdent of R when $R \geq 1$. Based on above considerations, we suggest the empirical formulas as $\Xi(\theta;x,R) \sim \Xi_0(x)\theta^2(1-\theta)^{-2x}$ and $\Sigma(\theta;x) \sim \Sigma_0(x)\theta(1-\theta)^{-(x+1)}$ for $x \geq 1, R \geq 1$, where $\Xi_0(x)$ and $\Sigma_0(x)$ can be determined by the best fitting at $\theta = 0.99$ (since in practice we usually have $\theta \to 1$).

Fig. 3 The number-average length $L_n(t)$ and weight-average length $L_w(t)$ of $PI_{264}$-*b*-$PFS_{48}$ micelles. Comparisons are made among TEM measurements by Guerin *et al.*[17] (black symbols), mass-action equations in Eq. 3 (red solid curves), and moment-closure equations in Eq. 10 (blue dashed lines). The parameters and initial values in calculation are given in accordance with the legend of Fig. 4. The zoomed-in plot adopts a log-scale of the sonication time to highlight the initial time-evolution region. The mismatch between the simulation and experimental data before $t = 10s$ mainly dues to the approximate Gaussian distribution we taken as a start.

Fig. 4 The weight length distribution of $PI_{264}$-*b*-$PFS_{48}$ micelles. Comparisons are made among TEM measurements by Guerin *et al.*[17] (black symbols), mass-action equations in Eq. 3 (red solid curves), moment-closure equations in Eq. 10 plus approximate distribution function in Eq. 8 (green dotted curves), and moment-closure equations in Eq. 10 plus approximate distribution function in Eq. 11 (blue dashed and dotted lines). In all calculations we take the same set of parameters $m_{tot} = 0.06 mg/mL$, $k_f^+ = 2 \times 10^{-9} s^{-1}$, $x = 2.6$, $R = 1$. The initial weight length distribution used in mass-action equations are taken as a Gaussian form $F_w(L) = \exp\left[-(L-L_w)^2/(2\sigma^2)\right]$ with $L_w = 2450nm$ and $\sigma = 148.63nm$. And the initial number concentration used in moment-closure equations are chosen as $P(0) = m_{tot}/1050$.

**FIGURES**

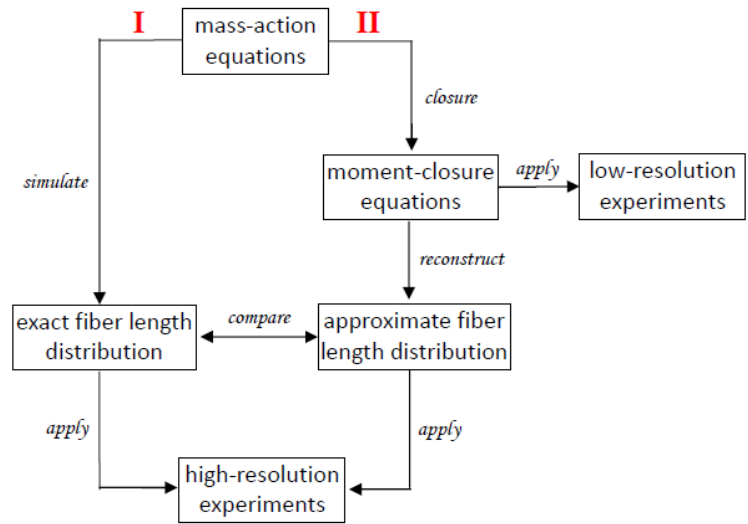

Fig. 1

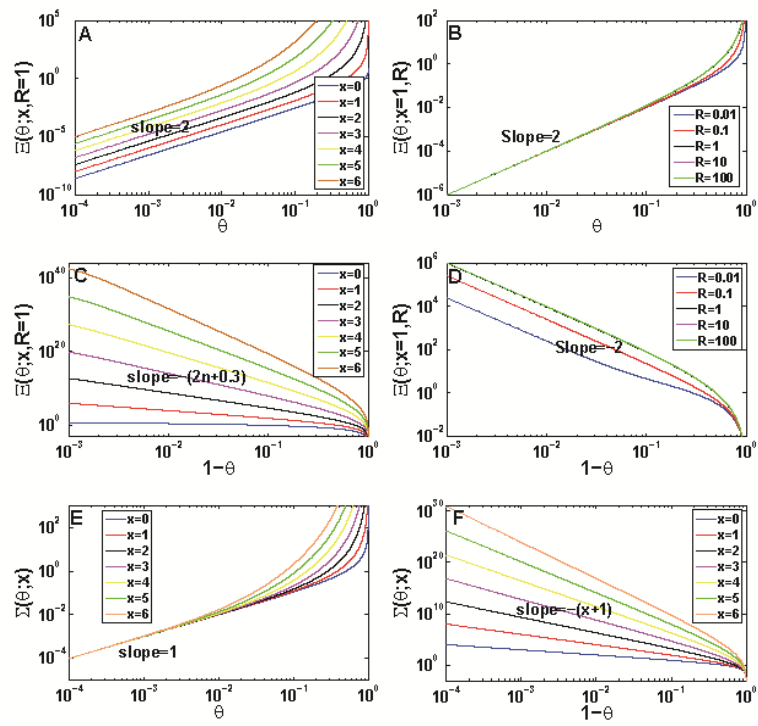

Fig. 2

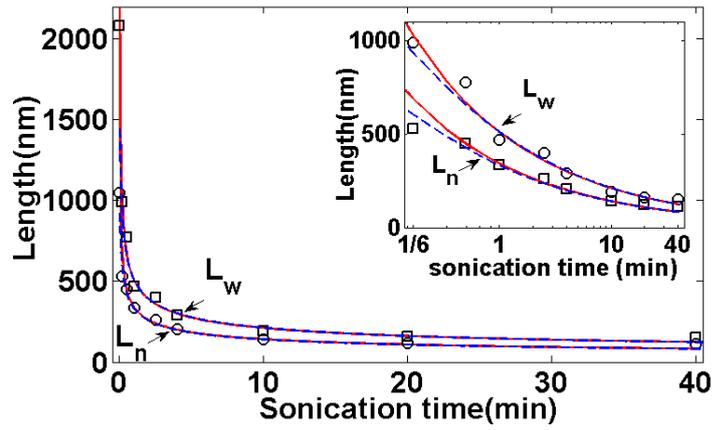

Fig. 3

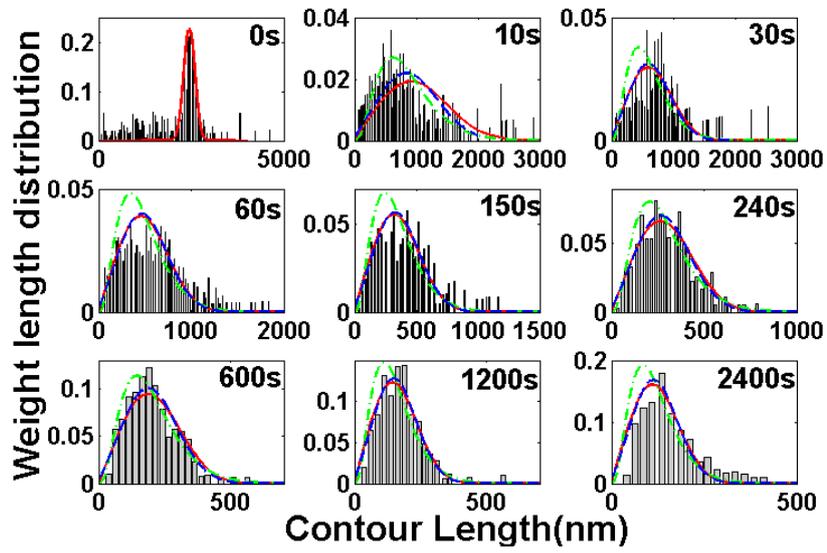

Fig. 4

## Supplementary Materials

## Hill's Model

Here we apply our method to a different model – Hill's model[1], in which the length-dependent reaction rate constants for fiber fragmentation and association are chosen as

$$\begin{cases} k_f^+(i,j) = k_f^+(ij)^{n-1}(i\ln j + j\ln i)/(i+j)^{n+1}, \\ k_f^-(i,j) = k_f^-(i\ln j + j\ln i)/[ij(i+j)], \end{cases} \quad (S1)$$

where $n$ represents degrees of freedom of a fiber moving in the solution.

Similarly, we introduce a free energy function $F$ as

$$\begin{aligned} F/(k_B T) = & \varepsilon_f \sum_{i=1}^{\infty}[A_i] + \varepsilon_m \sum_{i=1}^{\infty} i \cdot [A_i] \\ & + \sum_{i=1}^{\infty}([A_i]\ln[A_i] - [A_i]) - n\sum_{i=1}^{\infty} \ln i \cdot [A_i]. \end{aligned} \quad (S2)$$

The meaning of each term in above equation is the same as that in Eq. 4. And we further have $\varepsilon_f = -\ln(k_f^+/k_f^-) > 0$ according to the well-known relation between the Gibbs free energy change for a reaction and the equilibrium constant.

In the same way, we find the approximate fiber length distribution as

$$[A_i] = \frac{k_f^+}{k_f^-} i^n \cdot \exp\left[\lambda_1 + i(\lambda_2 + \varepsilon_m)\right], \quad (S3)$$

where the multipliers $\lambda_1$ and $\lambda_2$ can be determined from $P(t)$ and $M(t)$,

$$\begin{cases} \dfrac{k_f^+}{k_f^-} \sum_{i=1}^{\infty} i^n \cdot \exp\left[\lambda_1 + i(\lambda_2 + \varepsilon_m)\right] = P, \\ \dfrac{k_f^+}{k_f^-} \sum_{i=1}^{\infty} i^{n+1} \cdot \exp\left[\lambda_1 + i(\lambda_2 + \varepsilon_m)\right] = M. \end{cases} \quad (S4)$$

Substituting the distribution function in Eq. S3 into original mass-action equations (Eq. 2), we can derive the moment-closure equations for $P(t)$ as

$$\begin{cases} dP/dt = \left(k_f^+ \alpha - k_f^- \alpha^2\right) \Xi(\chi; n), \\ \alpha \Sigma(\chi; n) = P, \\ \alpha \Sigma(\chi; n+1) = M, \end{cases} \quad (S5)$$

where $\alpha = k_f^+ \exp(\lambda_1)/k_f^-$, $\chi = \exp(\lambda_2 + \varepsilon_m)$, and $M(t) = M(0)$ according to the conservation law of mass. The empirical functions

$$\Xi(\chi; n) = \sum_{i=1}^{\infty}\sum_{j=1}^{\infty} \frac{(ij)^{n-1}(i\ln j + j\ln i)}{i+j} \chi^{i+j} \sim \Xi_0(n)\chi^3 (1-\chi)^{-(2n+0.3)}$$

with $\Xi_0(n) \approx \exp(0.202n^2 + n - 3.3) + 0.6$ (see Figs. S1A and S1B), and

$$\Sigma(\chi; n) = \sum_{i=1}^{\infty} i^n \chi^i \sim \Sigma_0(n)\chi(1-\chi)^{-(n+1)}$$

with

$\Sigma_0(n) \approx \exp(0.092n^2 + 0.77n - 1.38) + 0.35$ (see Figs. 1E and 1F) are found through numerical experiments.

Actually when $n$ is a integer, the normalization conditions in Eq. S5 can be exactly solved, *i.e.*

$$\begin{cases} \Sigma(\chi;0) = \chi(1-\chi)^{-1}, \\ \Sigma(\chi;n) = \sum_{i=1}^{n} \dfrac{i! C_i^n \chi^i}{(1-\chi)^{i+1}}, \quad n \geq 1 \end{cases} \quad (S6)$$

where $C_i^j$ can be determined through the recursion formula $C_i^{j+1} = C_{i-1}^j + iC_i^j, C_1^j = C_j^j = 1$, or by following coefficient triangle

$$\begin{array}{cccccc} 1_1^1 & & & & & \\ 1_1^2 & 1_2^2 & & & & \\ 1_1^3 & 3_2^3 & 1_3^3 & & & \\ 1_1^4 & 7_2^4 & 6_3^4 & 1_4^4 & & \\ 1_1^5 & 15_2^5 & 25_3^5 & 10_4^5 & 1_5^5 & \\ 1_1^6 & 31_2^6 & 90_3^6 & 65_4^6 & 15_5^6 & 1_6^6 \\ \cdots & \cdots & \cdots & \cdots & \cdots & \cdots \end{array} \quad (S7)$$

Finally the fiber length distribution can be reconstructed according to Eqs. S3-S5. Quantitative comparisons with the sonication studies on $PI_{264}$-*b*-$PFS_{48}$ micelles performed by Guerin *et al.*[2] predict almost the same results (see Figs. S2 and S3), which further confirm the validity and generality of our method.

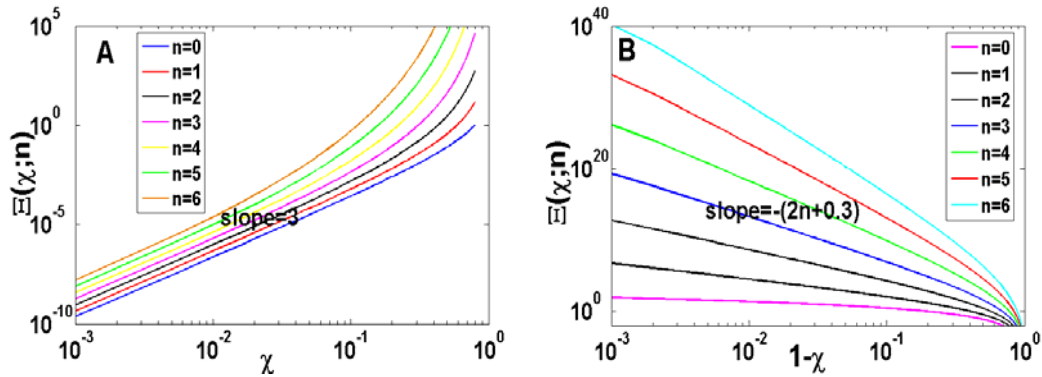

Fig. S1 Asympotic behaviors of functions (A-B) $\Xi(\chi;n)$. In the same way, two scaling regions are found, *i.e.* when $\chi \to 0$, $\Xi(\chi;n) \propto \chi^3$ ; when $\chi \to 1$,

$\Xi(\chi;n) \propto (1-\chi)^{-(2n+0.3)}$. Based on this consideration, we suggest the empirical formulas as $\Xi(\chi;n) \sim \Xi_0(n)\chi^3(1-\chi)^{-(2n+0.3)}$ for $n \geq 1$, where $\Xi_0(n)$ are determined by the best fitting at $\chi = 0.99$.

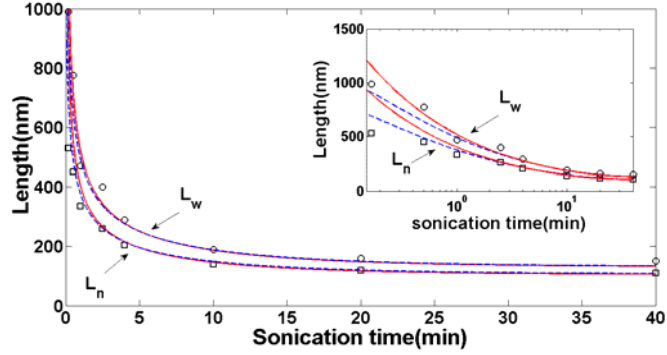

Fig. S2 The number-average length $L_n(t)$ and weight-average length $L_w(t)$ of PI$_{264}$-b-PFS$_{48}$ micelles. Comparisons are made among TEM measurements by Guerin et al.[2] (black symbols), mass-action equations in Eq. 3 (red solid curves), and moment-closure equations in Eq. S5 (blue dashed lines). The parameters and initial values in calculation are given in accordance with the legend of Fig. S3.

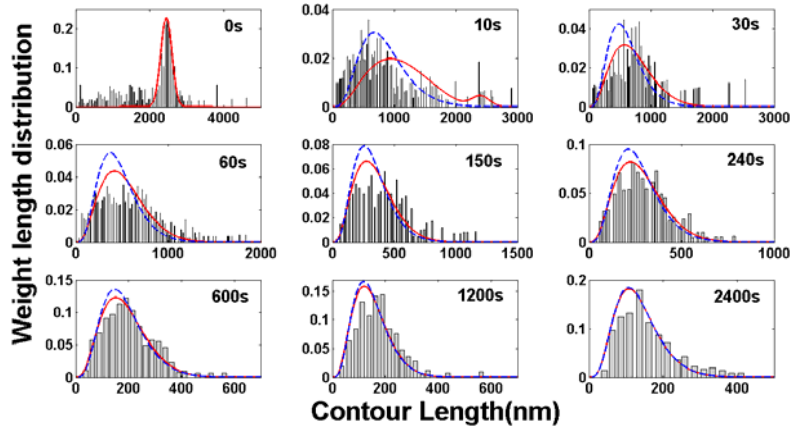

Fig. S3 The weight length distribution of PI$_{264}$-b-PFS$_{48}$ micelles. Comparisons are made among TEM measurements by Guerin et al.[2] (black symbols), mass-action equations in Eq. 3 (red solid curves), and moment-closure equations in Eq. S5 plus approximate distribution function in Eq. S3 (blue dashed lines). In all calculations we take the same set of parameters $m_{tot} = 0.06 mg/mL$, $k_f^+ = 2 \times 10^{-7} s^{-1}$, $k_f^- = 1 \times 10^3 (mg/mL)^{-1} s^{-1}$, $n = 3$. The initial weight length distribution used in mass-action equations are taken as a Gaussian form $F_w(L) = \exp\left[-(L-L_w)^2/(2\sigma^2)\right]$ with $L_w = 2450 nm$ and $\sigma = 148.63 nm$. And the initial number concentration used in moment-closure equations are chosen as $P(0) = m_{tot}/1050$.